\documentclass[showpacs,twocolumn,floatfix,aps]{revtex4}

\usepackage{graphicx}
\usepackage{dcolumn}
\usepackage{bm}
\usepackage{amsmath}
\usepackage{amssymb}
\usepackage{units}
\usepackage{url}
\usepackage{hyperref}
\usepackage{breakurl}

\begin{document}


\title{Thermodynamic properties of separable square-wave potentials}

\author{B. Kollmitzer}
 \affiliation{Institute of Solid State Physics, Graz University of Technology, Graz, Austria}
\author{P. Hadley}%
 \email{p.hadley@tugraz.at}
 \affiliation{Institute of Solid State Physics, Graz University of Technology, Graz, Austria}
\date{\today}

\begin{abstract} 
Exact analytic solutions to the Schr\"odinger equation for an electron moving 
in three dimensional potentials have been studied. These solutions can 
correspond to metals, semiconductors, or insulators. We show that there is an 
efficient method to calculate the electron density of states for this class 
of potentials. From the density of states, the temperature dependence of 
thermodynamic properties such as the chemical potential and the specific heat 
were determined. Ten thousand cubic separable potentials were considered. 
This data makes it possible to identify trends in how the form of the 
potential is related to the thermodynamic properties of a material. 
\end{abstract}

\maketitle

\section{Introduction}

To calculate the band structure of a crystal, some initial guess of the 
electron wave-functions is required. One of the rare cases where analytic 
solutions to the Schr\"odinger equation are known for a semiconductor or an 
insulator are separable potentials that can be written as the sum of three 
one-dimensional square-wave potentials. Such separable potentials were first 
considered by Kronig and Penney\cite{de_l._kronig_quantum_1931}. There is a 
simple analytic expression for the energy -- wave-number dispersion relation 
for the well-known problem of an electron moving in a one-dimensional 
square-wave potential. Here we show that there is also an analytic expression for 
the density of states for the one-dimensional problem. These results can be 
simply combined to produce the dispersion relations and density of states of 
separable three dimensional potentials. A limiting case of this class of 
potentials is the widely used free electron model which occurs when the 
amplitude of the square-wave is zero. As the amplitude of the potential 
increases, the potentials can correspond to metals, semiconductors, or 
insulators. Normally determining the electronic density of states from the 
band structure is a computationally intensive process that requires sampling 
many points of the Brillouin zone. For the separable square-wave potentials, 
the electronic density of states is however easily calculable as a 
convolution of three analytically known functions. This makes it easy to 
calculate the thermodynamic properties corresponding to this class of 
potentials. From the density of states we can calculate the temperature 
dependence of thermodynamic quantities such as the chemical potential, the 
internal energy, the specific heat, the entropy, and the Helmholtz free 
energy. Half a million plots of the thermodynamic properties of separable 
potentials are available as online supplementary material\cite{supp}. The 
collective data set allows us to 
consider how an increase in the amplitude of the potential causes a 
transition from a metal to an insulator. There are cases where increasing the 
amplitude of the potential causes a transition from a metal to an insulator, 
then back to a metal and then finally back to an insulator. 

While band gaps appear in the one-dimensional problem for an arbitrarily small 
amplitude of the periodic potential, a finite amplitude is needed for the creation 
of band gaps in three dimensions. In section \ref{sec:3} we quantify how large 
the amplitude of the potential must be for bands to be formed for the specific 
class of three-dimensional potentials we have considered. This minimum amplitude is,
\begin{equation}
V = 0.9\frac{{\pi ^2 \hbar ^2 }}{{ma^2 }}.
\label{eq:1}
\end{equation}
Here $V$ is the amplitude of the potential, $\hbar$ is the reduced Planck's constant, $m$ 
is the mass of an electron, and $a$ is the lattice constant. The amplitude of the potential 
necessary for the formation of bands depends strongly on the lattice constant.

\section{Solutions of the Schr\"odinger equation for the Kronig-Penney potential}

\begin{figure}
\includegraphics{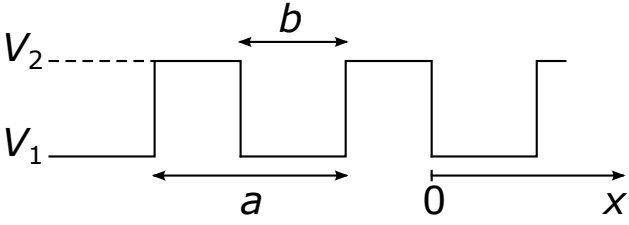}
\caption{\label{fig:1} One-dimensional square-wave potential.}
\end{figure}

This section reviews the solutions to the Kronig-Penney 
model\cite{mcquarrie_1996,szumlowicz_1997} and derives an expression for 
the one-dimensional density of states which will be used in section \ref{sec:3} to construct 
the dispersion relations and densities of states for three-dimensional separable potentials. 
Figure \ref{fig:1} shows the one-dimensional potential that was first considered by Kronig 
and Penney\cite{de_l._kronig_quantum_1931}. 

Because of the translational symmetry of the potential, the eigenfunctions of the Hamiltonian 
are simultaneously eigenfunctions of the translation operator. The eigenfunctions of the 
translation operator $T$ can be readily constructed from any two linearly independent solutions 
of the one-dimensional Schr\"odinger equation. A convenient choice is,
\begin{equation}
\psi _1 (0) = 1,{\rm{      }}\frac{{d\psi _1 }}{{dx}}(0) = 0,{\rm{     }}\psi _2 (0) = 0,{\rm{      }}\frac{{d\psi _2 }}{{dx}}(0) = 1.
\label{eq:2}
\end{equation}

The solutions in region 1 ($0 < x < b$) are,
\begin{equation}
\psi _1 (x) = \cos (k_1 x),{\rm{                   }}\psi _2 (x) = \frac{{\sin (k_1 x)}}{{k_1 }},
\label{eq:3}
\end{equation}
while the solutions in region 2 ($b < x < a$) are\cite{merzbacher_1961},
\begin{subequations}
\label{eq:4}
\begin{align}
 \psi _1 (x) = \cos (k_2 (x - b))\cos (k_1 b) \nonumber\\
 - \frac{k_1}{k_2} \sin (k_2 (x - b))\sin (k_1 b), 
\end{align}
\begin{align}
 \psi _2 (x) = \frac{{1}}{{k_1 }} \cos (k_2 (x - b))\sin (k_1 b)\nonumber\\
 + \frac{{1}}{{k_2 }} \sin (k_2 (x - b))\cos (k_1 b). 
\end{align}
\end{subequations}
Here $k_r = \sqrt {2m\left( {E - V_r } \right)/\hbar ^2 }$; $r = 1,2$ is the label of the 
region, and $E$ is the energy. For energies where $k_r$ is imaginary, the solutions are still real 
since $\cos(i\theta) = \cosh(\theta)$ and $\sin(i\theta) = i \sinh(\theta)$. 

Any other solution can be written as a linear combination of $\psi_1(x)$ and $\psi_2(x)$. In 
particular, $\psi_1(x + a)$ and $\psi_2(x + a)$ can be written in terms of $\psi_1(x)$ and 
$\psi_2(x)$. These solutions are related to each other by the matrix representation of the 
translation operator\cite{magnus_hills_1966}.
\begin{equation}
\left[ {\begin{array}{*{20}c}
   {\psi _1 \left( {x + a} \right)}  \\
   {\psi _2 \left( {x + a} \right)}  \\
\end{array}} \right] = \left[ {\begin{array}{*{20}c}
   {T_{11} } & {T_{12} }  \\
   {T_{21} } & {T_{22} }  \\
\end{array}} \right]\left[ {\begin{array}{*{20}c}
   {\psi _1 \left( x \right)}  \\
   {\psi _2 \left( x \right)}  \\
\end{array}} \right].
\label{eq:5}
\end{equation}
The elements of the translation matrix can be determined by evaluating Eq. \eqref{eq:5} and 
its derivative at $x = 0$.
\begin{equation}
\left[ {\begin{array}{*{20}c}
   {\psi _1 \left( {x + a} \right)}  \\
   {\psi _2 \left( {x + a} \right)}  \\
\end{array}} \right] = \left[ {\begin{array}{*{20}c}
   {\psi _1 \left( a \right)} & {\frac{{d\psi _1 }}{{dx}}\left( a \right)}  \\
   {\psi _2 \left( a \right)} & {\frac{{d\psi _2 }}{{dx}}\left( a \right)}  \\
\end{array}} \right]\left[ {\begin{array}{*{20}c}
   {\psi _1 \left( x \right)}  \\
   {\psi _2 \left( x \right)}  \\
\end{array}} \right].
\label{eq:6}
\end{equation}
The eigenfunctions and eigenvalues $\lambda$ of the translation operator are, 
\begin{align}
\psi _ \pm  \left( x \right) = \frac{{2\psi _2 (a)}}{{\frac{{d\psi _2 (a)}}{{dx}} - \psi _1 (a) \pm \delta }}\psi _1 (x) + \psi _2 (x),\nonumber\\
{\rm{         }}\lambda _ \pm = \frac{1}{2}\left( {\alpha  \pm \delta } \right),
\label{eq:7}
\end{align}
where $\delta  = \sqrt {\alpha ^2  - 4} {\rm{  }}$ and 
\begin{align}
\alpha  = \psi _1 (a) + \frac{{d\psi _2 (a)}}{{dx}} = 2\cos \left( {k_2 \left( {a - b} \right)} \right)\cos \left( {k_1 b} \right) \nonumber\\
- \left( {\frac{{k_2 }}{{k_1 }} + \frac{{k_1 }}{{k_2 }}} \right)\sin \left( {k_2 \left( {a - b} \right)} \right)\sin \left( {k_1 b} \right).
\label{eq:8}
\end{align}

If periodic boundary conditions are used for a potential with $N$ unit cells, then 
applying the translation operator $N$ times brings the function back to its original 
position 
\begin{equation}
T^N \psi \left( x \right) = \psi \left( {x + Na} \right) = \lambda ^N \psi \left( x \right) = \psi \left( x \right).
\label{eq:9}
\end{equation}
The eigenvalues of the translation operator are therefore the solutions to the equation 
$\lambda^N = 1$. These solutions are,
\begin{equation}
\lambda _j  = \exp \left( \frac{i2\pi j}{N} \right) = \exp \left( \frac{i2\pi aj}{L} \right) = \exp \left( {ik_j a} \right),
\label{eq:10}
\end{equation}
where $j$ is an integer between $-N/2$ and $N/2$, $L = Na$ is the length of the crystal, 
and $k_j = 2\pi j/L$ are the allowed $k$ values in the first Brillouin zone. The dispersion 
relation can be determined by first calculating $\alpha$ for a specific energy and then 
solving Eqs. \eqref{eq:7} and \eqref{eq:10} for the wave-number\cite{bender_1999}, 
\begin{equation}
k =  \pm \frac{1}{a}\tan ^{ - 1} \left( {\frac{{\sqrt {4 - \alpha ^2 } }}{\alpha }} \right).
\label{eq:11}
\end{equation}
\begin{figure*}
\includegraphics{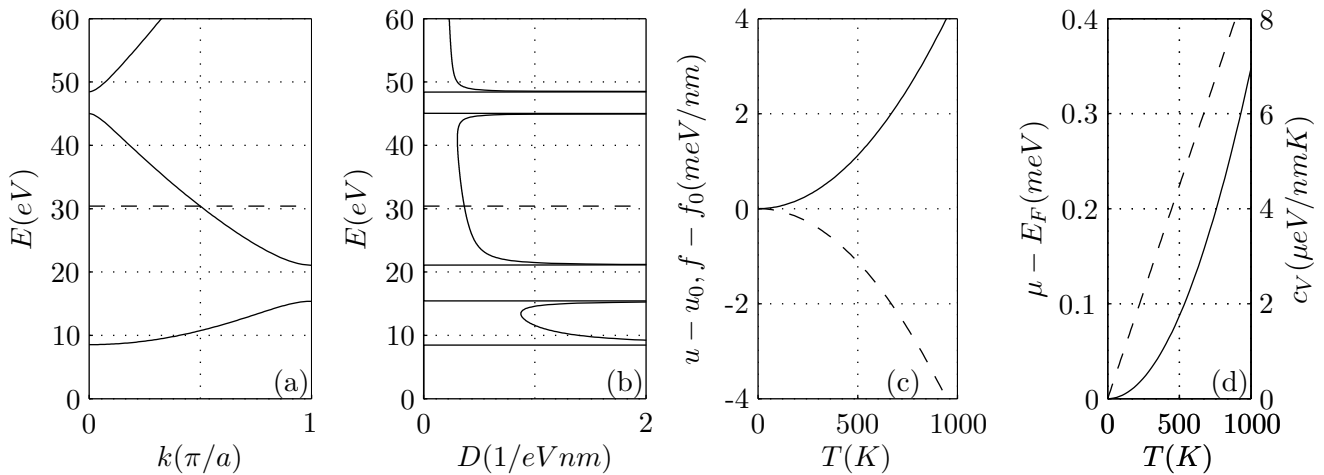}
\caption{\label{fig:2}(a) The energy -- wave-number dispersion relation. The dashed 
line is the Fermi energy. (b) The density of states. (c) The internal energy 
density (solid line) and Helmholtz free energy density (dashed line). (d) The 
chemical potential (solid line) and the specific heat (dashed line). All of the 
plots were drawn for a square-wave potential with the parameters: $V_1 = \unit[0]{eV}$, 
$V_2 = \unit[12.5]{eV}$, $a = \unit[2 \times 10^{-10}]{m}$, $b = \unit[5 \times 10^{-11}]{m}$, 
and an electron density of $n = \unit[3]{}${electrons/primitive cell}.}
\end{figure*}

The dispersion relation can be used to determine the density of states which is 
needed to calculate the thermodynamic properties of a system of noninteracting 
electrons\cite{lin_2003}. The one-dimensional density of states in $k$ space is $D(k) = 2/\pi$ 
and thus the density of states in energy is,
\begin{equation}
D(E) = 
\begin{cases}
 \frac{2}{\pi }\frac{{dk}}{{dE}} = \frac{2}{\pi }\frac{{dk}}{{d\alpha }}\frac{{d\alpha }}{{dE}} & \left| \alpha  \right| < 2 \\
 0 & \left| \alpha  \right| > 2
\end{cases}
\label{eq:12}
\end{equation}
where 
\begin{equation*}
\frac{{dk}}{{d\alpha }} = \frac{1}{{a\sqrt {4 - \alpha ^2 } }},
\end{equation*}
and
\begin{widetext}
\begin{align}
\frac{{d\alpha }}{{dE}} =  - \frac{m}{{\hbar ^2 }}\left( {\frac{{2b}}{{k_1 }} + \left( {\frac{{k_2 }}{{k_1 }} + \frac{{k_1 }}{{k_2 }}} \right)\frac{{\left( {a - b} \right)}}{{k_2 }}} \right)\cos \left( {k_2 \left( {a - b} \right)} \right) \sin \left( {k_1 b} \right) - \frac{m}{{\hbar ^2 }}\left( {2\frac{{\left( {a - b} \right)}}{{k_2 }} + \left( {\frac{{k_2 }}{{k_1 }} + \frac{{k_1 }}{{k_2 }}} \right)\frac{b}{{k_1 }}} \right)  \nonumber\\
\times \sin \left( {k_2 \left( {a - b} \right)} \right) \cos \left( {k_1 b} \right) + \frac{m}{{\hbar ^2 }}\left( {\left( {\frac{{k_2 }}{{k_1^2 }} - \frac{1}{{k_2 }}} \right)\frac{1}{{k_1 }} + \left( {\frac{{k_1 }}{{k_2^2 }} - \frac{1}{{k_1 }}} \right)\frac{1}{{k_2 }}} \right) \sin \left( {k_2 \left( {a - b} \right)} \right)\sin \left( {k_1 b} \right). 
\label{eq:13}
\end{align}
\end{widetext}

Thermodynamic properties such as the chemical potential, the internal energy, 
the specific heat, the entropy, or the Helmholtz free energy of a system of 
noninteracting fermions can be calculated from the electron density $n$, the density 
of states $D(E)$, and the temperature $T$. 

The chemical potential $\mu$ is implicitly defined by the relation, 
\begin{equation}
n = \int\limits_{ - \infty }^\infty  {D(E)F(E)dE} ,
\label{eq:14}
\end{equation}
where $F(E)$ is the Fermi function,
\begin{equation}
F(E) = \frac{1}{{\exp \left( {\frac{{E - \mu }}{{k_B T}}} \right) + 1}}.
\label{eq:15}
\end{equation}
Here $k_B$ is Boltzmann's constant. Once the chemical potential has been determined, 
it can be used to calculate the internal energy density $u$ and the Helmholtz free 
energy density $f$. 
\begin{equation}
u = \int\limits_{ - \infty }^\infty  {ED(E)F(E)dE} 
\label{eq:16}
\end{equation}
\begin{equation}
f = \mu n - k_B T\int\limits_{ - \infty }^\infty  {D(E)\ln \left( {1 + \exp \left( {\frac{{\mu  - E}}{{k_B T}}} \right)} \right)dE} 
\label{eq:17}
\end{equation}
Finally, the specific heat and the entropy are given by the partial derivatives,
\begin{equation}
c_v  = \left. {\frac{{\partial u}}{{\partial T}}} \right|_{N,V} \text{ and }
s = - \left. {\frac{{\partial f}}{{\partial T}}} \right|_{N,V}.
\label{eq:18}
\end{equation}

The band structure of a one dimensional potential and the corresponding thermodynamic 
properties are plotted in Fig. \ref{fig:2}. 

\section{Three-dimensional separable potentials} \label{sec:3}

In this section, 5 of the calculated band structures that are available in 
the supplementary material are presented. This illustrates the variety of results 
that can be obtained with this simple model. The five examples are a free electron 
gas, a metal where the charge carriers at the Fermi surface are electron-like, a 
metal where the charge carriers at the Fermi surface are hole-like, a direct band 
gap semiconductor, and an indirect band gap semiconductor. 

The dispersion relation and density of states for any three-dimensional potential 
of the form
\begin{equation}
U_{3d}(x,y,z) = U_x(x) + U_y(y) + U_z(z)
\label{eq:19}
\end{equation}
are easily calculated from the one-dimensional results\cite{berezin_1986}. 
The energy of an electron in a three-dimensional separable potential is the sum of the 
energies of the constituent one-dimensional potentials. The three-dimensional density of 
states is the convolution of the three one-dimensional densities of states. Once the 
three-dimensional density of states is known, the thermodynamic quantities can be 
calculated as outlined above. 

The one-dimensional bands can be indexed by integers, 1 corresponding to the band 
with the lowest energy, 2 to the band with the next lowest energy, etc. The three-dimensional 
bands can then be indexed by the three integers that correspond to the 
one-dimensional bands that make up the three-dimensional band. The three dimensional 
band with the lowest energy is the 111 band. For cubic crystals, the next three 
bands 211, 121, and 112 are degenerate in energy. 

\begin{figure*}
\includegraphics{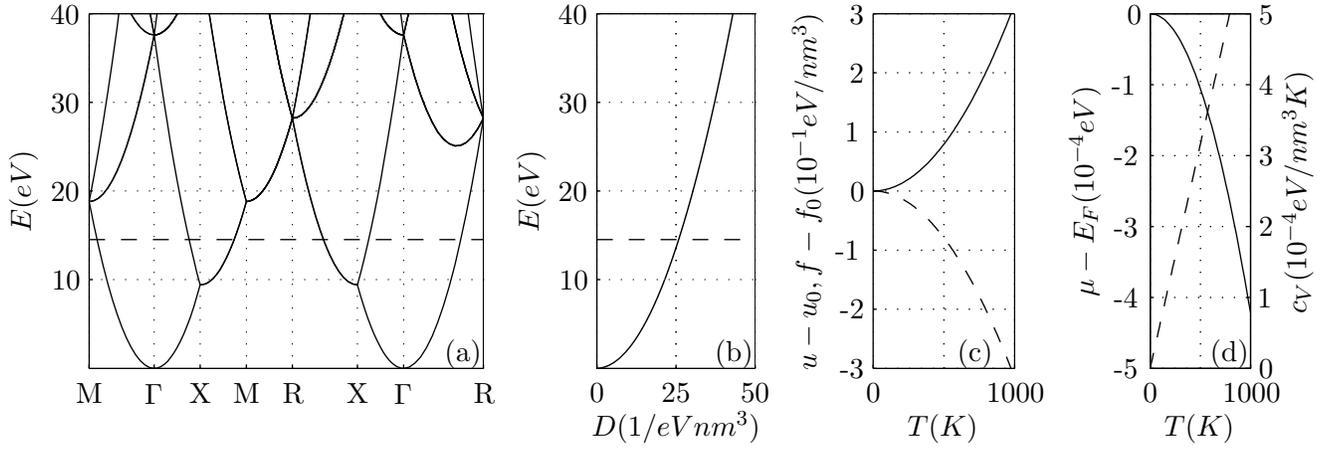}
\caption{\label{fig:3}The band structure and thermodynamic properties of a free electron 
gas. (a) The energy-momentum dispersion relation. The dashed line is the Fermi energy. 
(b) The density of states. (c) The internal energy density (solid line) and the Helmholtz 
free energy density (dashed line). (d) The chemical potential (solid line) and the 
specific heat (dashed line). All plots are for a cubic potential with the parameters: 
$V_1 = \unit[0]{eV}$, $V_2 = \unit[0]{eV}$, $a = \unit[0.2]{nm}$, $b = \unit[0.09]{nm}$, 
and an electron density of $n = \unit[2]{}${electrons/primitive cell}.}
\end{figure*}

Figure \ref{fig:3} shows the band structure, the density of states, and corresponding 
thermodynamic properties for a constant potential that corresponds to a free electron 
gas. The band structure in Fig. \ref{fig:3}a is plotted along a path in the Brillouin 
zone going from the M point (0.5, 0.5, 0) through the points $\Gamma$ (0, 0, 0), 
X(0.5, 0, 0), M, R (0.5, 0.5, 0.5), X, $\Gamma$, ending at the point R. All of the 
standard results for a free electron gas are reproduced by the numerical calculation. 
The dispersion relation is parabolic; the density of states increases with the square 
root of the energy and the specific heat is a linear function of the temperature. 

\begin{figure*}
\includegraphics{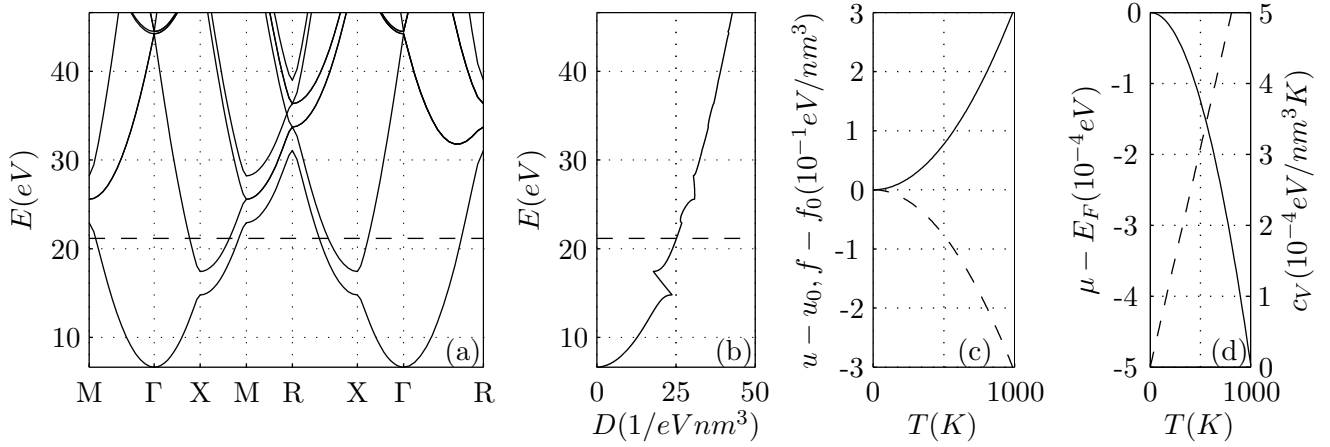}
\caption{\label{fig:4}The band structure and thermodynamic properties for a cubic potential 
$V_1 = \unit[0]{eV}$, $V_2 = \unit[4.2]{eV}$, $a = \unit[0.2]{nm}$, $b = \unit[0.09]{nm}$ 
and $n = \unit[2]{}$electrons/primitive cell. (a)-(d) as in Fig. \ref{fig:3}.}
\end{figure*}

\begin{figure*}
\includegraphics{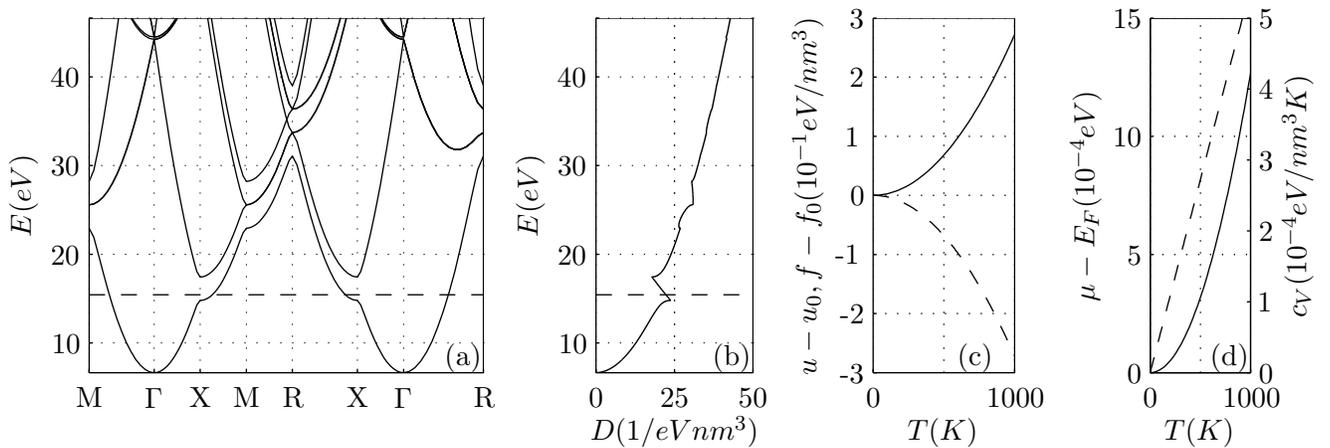}
\caption{\label{fig:5}The band structure and thermodynamic properties for a cubic potential 
$V_1 = \unit[0]{eV}$, $V_2 = \unit[4.2]{eV}$, $a = \unit[0.2]{nm}$, $b = \unit[0.09]{nm}$ 
and $n = \unit[1]{}$electrons/primitive cell. (a)-(d) as in Fig. \ref{fig:3}.}
\end{figure*}

Figures \ref{fig:4} and \ref{fig:5} show the results for a potential with an amplitude 
of $\unit[4.2]{eV}$, $a = \unit[0.2]{nm}$, and $b = \unit[0.09]{nm}$. Gaps open in the 
dispersion relation at the Brillouin zone boundaries and some kinks appear in the density 
of states. These kinks make both the analytical and numerical evaluation of the 
thermodynamic properties difficult. Both Figs. \ref{fig:4} and \ref{fig:5} correspond 
to metals but in Fig. \ref{fig:4} the electron density is $\unit[2]{}$electrons/primitive 
cell and the density of states is increasing at the Fermi energy while in Fig. \ref{fig:5} the 
electron density is $\unit[1]{}$electron/unit cell and the density of states is decreasing 
at the Fermi energy. The material in Fig. \ref{fig:4} has a Fermi surface that consists 
of electron-like states and the chemical potential decreases with increasing temperature 
as it does for the free electron gas of Fig. \ref{fig:3}. The material in Fig. 
\ref{fig:5} has a Fermi surface that consists of hole-like states and consequently the 
chemical potential increases with increasing temperature. 

\begin{figure*}
\includegraphics{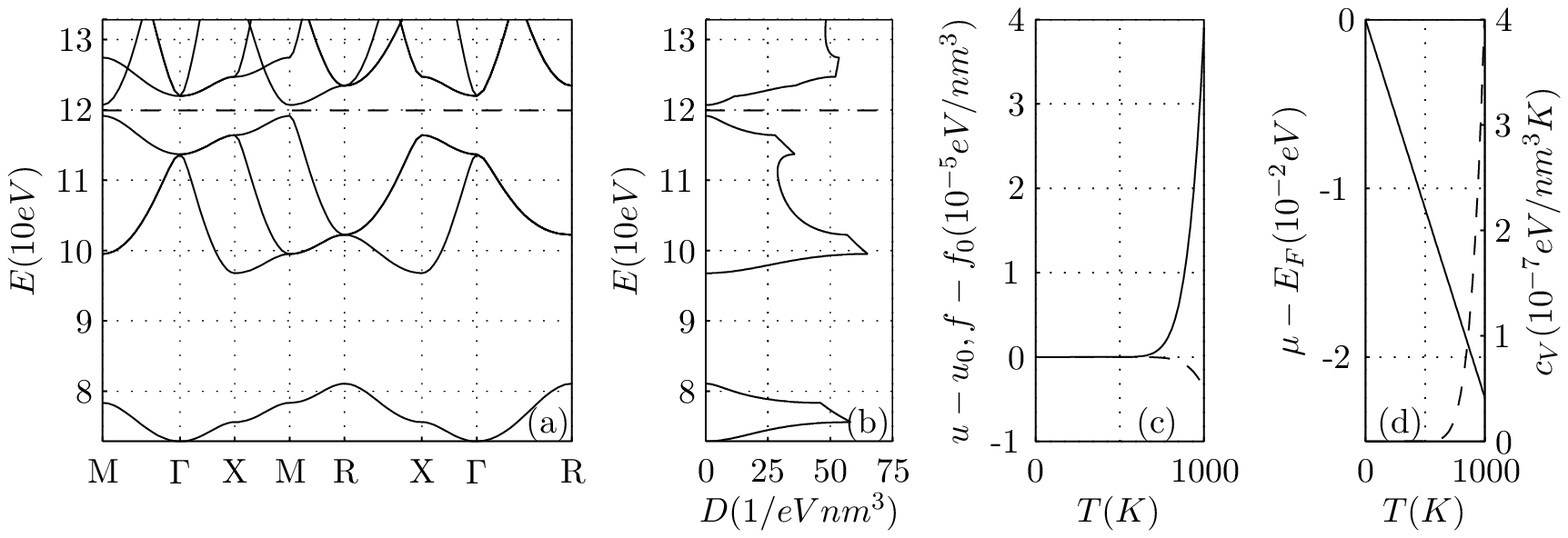}
\caption{\label{fig:6}The band structure and thermodynamic properties for a cubic potential 
$V_1 = \unit[0]{eV}$, $V_2 = \unit[8.8]{eV}$, $a = \unit[0.4]{nm}$, $b = \unit[0.21]{nm}$ 
and $n = \unit[8]{}$electrons/primitive cell. This band structure corresponds to a direct 
band gap semiconductor. (a)-(d) as in Fig. \ref{fig:3}.}
\end{figure*}

Figure \ref{fig:6} shows the band structure for a material with two band gaps. 
The lower band gap is indirect with a band gap energy of $\unit[3.9]{eV}$. The 
upper band is direct with a band gap energy of $\unit[0.4]{eV}$. If the material 
shown in Fig. \ref{fig:6} had $\unit[2]{}$electrons/primitive cell, the lowest 
band would be completely filled and it would be an insulator. For an electron 
density of $\unit[8]{}$electrons/primitive cell, the Fermi energy lies in the 
second band gap (illustrated by the dashed line in Fig. \ref{fig:6}a. The 
thermodynamic properties in Figs. \ref{fig:6}c and \ref{fig:6}d were calculated 
assuming an electron density of $\unit[8]{}$electrons/primitive cell. Since this 
material is a semiconductor, the electronic contribution to the specific heat is 
negligible at low temperature and increases exponentially at high temperatures. 
The chemical potential of a semiconductor is a linear function of the temperature. 

\begin{figure*}
\includegraphics{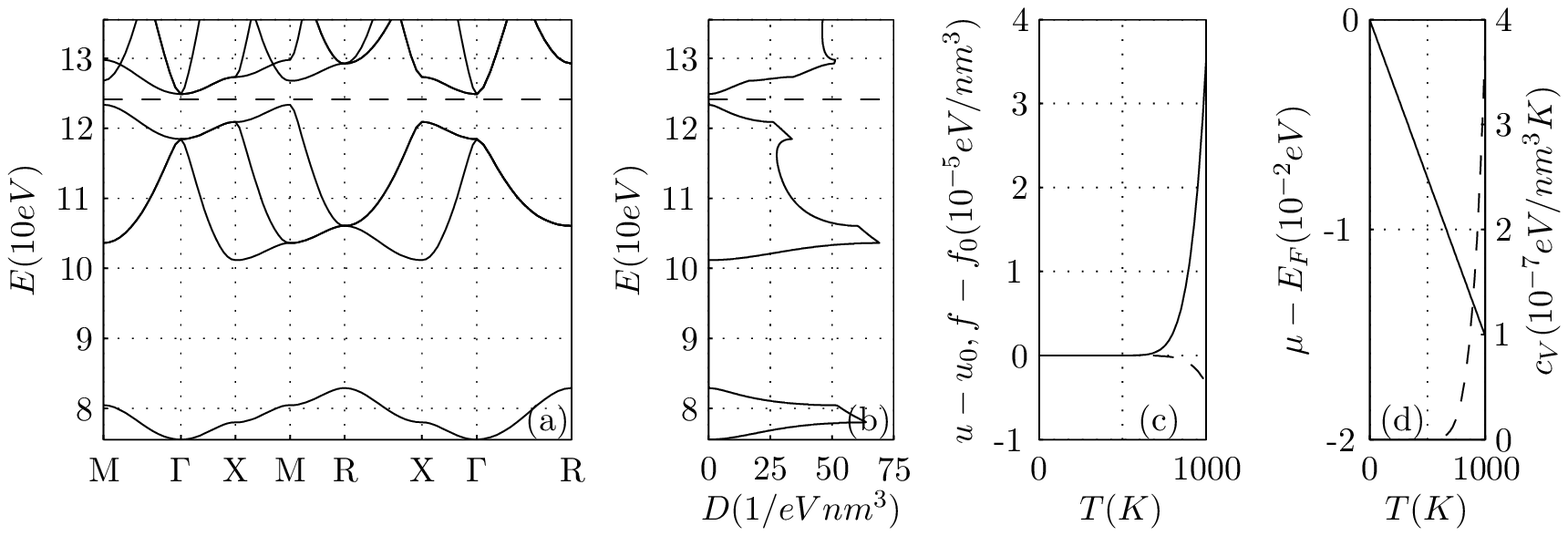}
\caption{\label{fig:7}The band structure and thermodynamic properties for a cubic potential 
$V_1 = \unit[0]{eV}$, $V_2 = \unit[9.5]{eV}$, $a = \unit[0.4]{nm}$, $b = \unit[0.19]{nm}$ 
and $n = \unit[8]{}$electrons/primitive cell. This band structure corresponds to an indirect 
band gap semiconductor. (a)-(d) as in Fig. \ref{fig:3}.}
\end{figure*}

Figure \ref{fig:7} shows the band structure for a potential where the 311 band has 
moved down lower than the 221 band. In this case the lower band is indirect with 
a band gap of $\unit[4.7]{eV}$ and the upper band gap is also indirect with a 
band gap energy of $\unit[0.5]{eV}$. For an electron density of 
$\unit[8]{}$electrons/primitive cell, the Fermi energy lies in the second band 
gap where the dashed line is drawn and the material is an indirect band gap 
semiconductor. 

The dispersion relations, densities of states, and thermodynamic properties like 
those shown in Fig. \ref{fig:3}-\ref{fig:7} were calculated for 10000 cubic potentials. 
For these calculations, it is convenient to normalize the Schr\"odinger equation 
so that length is measured in terms of the lattice constant $a$. When this is done, 
energies are measured in units of $\pi^2\hbar^2 / 2ma^2$ and there are only two 
independent parameters in the problem, which can be taken to be $\tilde{b} = b/a$ 
and $\tilde{V} = 2ma^2V / \pi^2\hbar^2$. Here $V = V_2 - V_1$. The calculations 
were performed for normalized parameters in the range $\tilde{b} = \left( 0,1 \right)$ 
and $\tilde{V} = \left( 0,27 \right)$. It becomes more difficult to perform the 
numerical calculations of the thermodynamic properties for large values of $\tilde{V}$. 
The upper limit of $\tilde{V} = 27$ corresponds to the amplitude where our 
calculations of the thermodynamic properties become unreliable. 

\begin{figure*}
\centering
\includegraphics{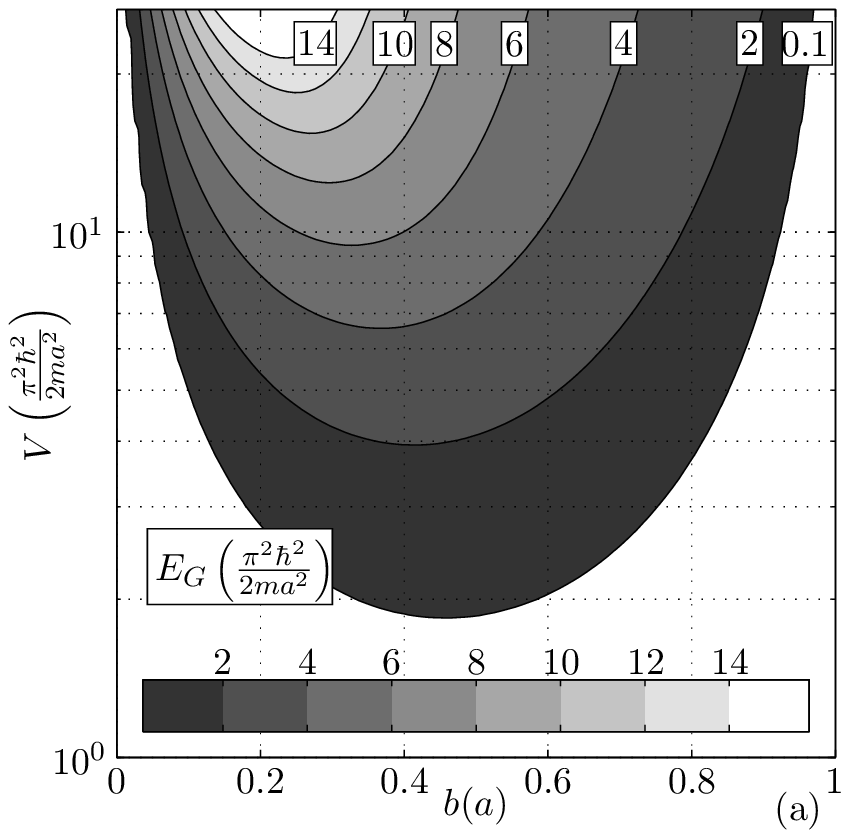}\hfill
\includegraphics{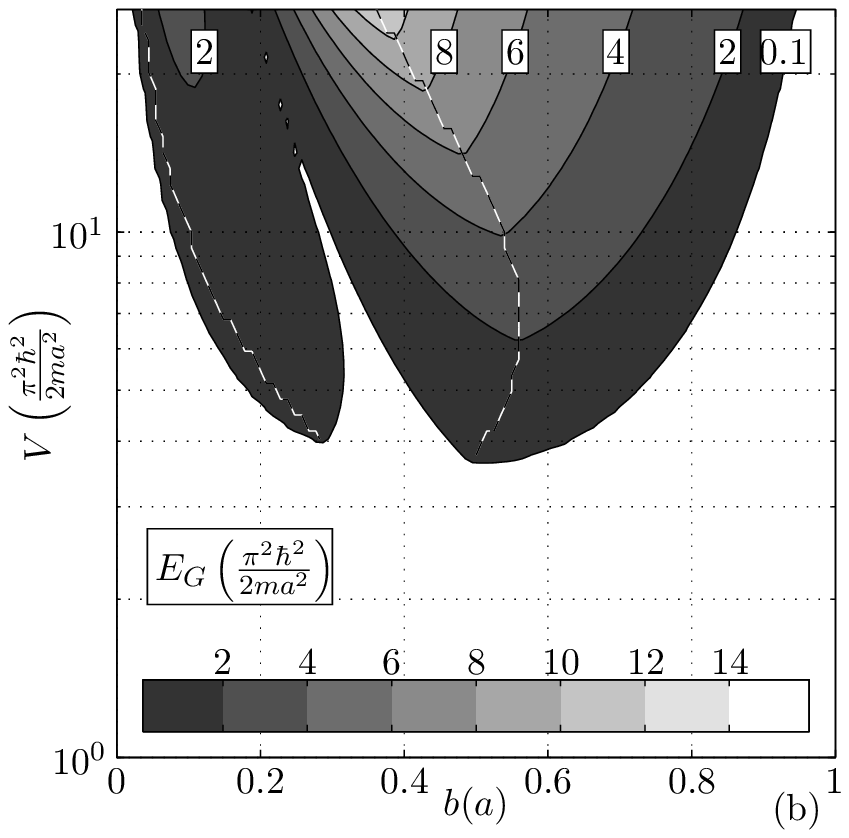}
\caption{\label{fig:8} (a) The band gap of the lowest gap between band 111 and the 
degenerate bands 112, 121, and 211 in units of $\pi^2\hbar^2 / 2ma^2$. This band 
gap is always indirect. (b) The second band gap. The dashed lines in (b) indicate 
the transitions from direct to indirect band gaps. The indirect band gaps are in 
between the two dashed lines. Note that there is a reentrant regime at $\tilde{b} = 0.3$
where an increase in the amplitude $V$ causes a transition from metal to insulator, 
then back to a metal and finally to an insulator.}
\end{figure*}

Figure \ref{fig:8} shows the sizes of the first two band gaps that occur. The 
three-dimensional potential must have a finite amplitude for there to be a band gap. 

\section{Thermodynamic properties of the metals} \label{sec:4}
The band structures of the potentials that were studied always correspond to 
metals when there is no band gap. This region includes the free electron model 
at $\tilde{V} = 0$ as well as $\delta$-function potentials in the regions 
near $\tilde{b} = 0$ and $\tilde{b} = 1$. When band gaps occur, the situation 
is more complicated. The potentials can correspond to metals, semiconductors, 
or insulators depending on the electron density. We compared the thermodynamic 
properties of these metals to results obtained by using the Sommerfeld expansion. 
Sommerfeld showed that the electronic contribution to the thermodynamic 
properties of metals can be approximated in terms of just two quantities: the 
density of states at the Fermi energy $D(E_F)$ and the derivative of the density 
of states at the Fermi energy\cite{sommerfeld_zur_1928}
\begin{equation}
D'(E_F ) = \left. {\frac{{dD(E)}}{{dE}}} \right|_{E_F }.
\label{eq:20}
\end{equation}

To lowest order in the temperature $T$, the Sommerfeld expression for the 
chemical potential, the internal energy, specific heat, entropy, and Helmholtz 
free energy are,
\begin{equation}
\mu  \approx E_F  - \frac{{\pi ^2 }}{6}\left( {k_B T} \right)^2 \frac{{D'(E_F )}}{{D(E_F )}},
\label{eq:21}
\end{equation}
\begin{equation}
u \approx u(T = 0) + \frac{{\pi ^2 D(E_F )}}{6}\left( {k_B T} \right)^2,
\label{eq:22}
\end{equation}
\begin{equation}
c_v  \approx \frac{{\pi ^2 D(E_F )}}{3}k_B^2 T,
\label{eq:23}
\end{equation}
\begin{equation}
s \approx \frac{{\pi ^2 D(E_F )}}{3}k_B^2 T,
\label{eq:24}
\end{equation}
\begin{equation}
f \approx u(T = 0) - \frac{{\pi ^2 D(E_F )}}{6}\left( {k_B T} \right)^2.
\label{eq:25}
\end{equation}

\begin{figure*}
\includegraphics{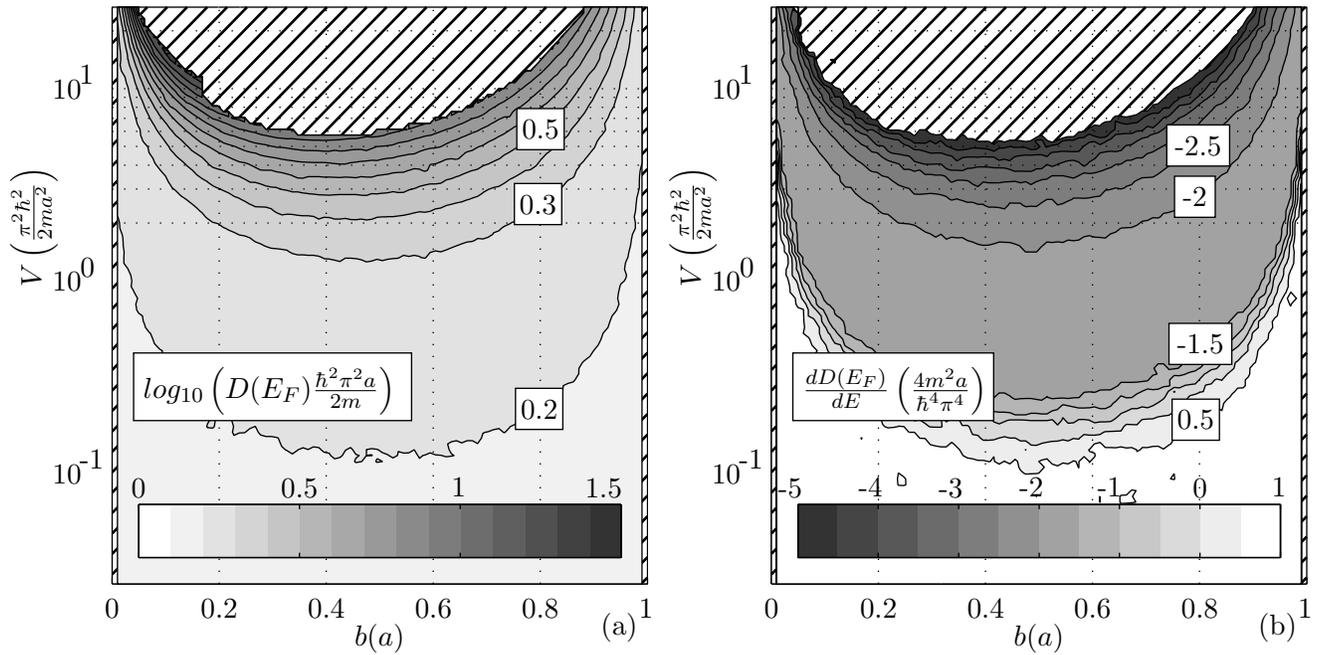}
\caption{\label{fig:9} The density of states at the Fermi energy and the 
derivative of the density of states at the Fermi energy for an electron density 
of $n = \unit[1]{}$electrons/primitive cell. It was not possible to determine 
the density of states at the Fermi energy reliably in the crosshatched regions. 
Similar plots for electron densities $\unit[2\text{--}10]{}$electrons/primitive cell are 
available in the supplementary information online.}
\end{figure*}

Figure \ref{fig:9} shows the value of the density of states at the Fermi energy 
and its derivative for an electron density of $n = \unit[1]{}$electron/primitive 
cell. This figure makes it possible to estimate the thermodynamic properties of 
the metals by substituting the values for $D(E_F)$ and $D'(E_F)$ into 
Eqs. \eqref{eq:21}--\eqref{eq:25}. 

The Sommerfeld expansion assumes that the density of states is a smoothly varying 
function in an energy range about $k_BT$ wide in the vicinity of the Fermi energy. 
However, the density of states must contain kinks known as Van Hove 
singularities\cite{ziman_principles_1969}. If the Fermi energy is within an energy $k_BT$ of one 
of the kinks, then the assumptions of the Sommerfeld theory are invalid. The numerical 
simulations in the supplementary material show that the Sommerfeld approximation is 
valid in the majority of cases where there are no band gaps. When gaps are present, 
the bands are narrow and it is more likely that the Sommerfeld theory fails. In the 
case of extremely narrow bands it becomes difficult to determine the temperature 
dependence of the thermodynamic properties reliably because of the discontinuities 
in the density of states.

\section{Thermodynamic properties of the semiconductors and insulators} \label{sec:5}

If the Fermi energy falls in a band gap so that the material is a semiconductor or an 
insulator, the density of states at the Fermi energy and its derivative are not 
suitable for describing the thermodynamic quantities. Instead, it is 
common to specify the band gap and the effective masses of the electrons and holes. 

For cubic crystals, the first gap appears above the 111 band and beneath the next 
three bands (211, 121, and 112) which are degenerate due to symmetry reasons. Above 
these three degenerate bands, a second energy gap sometimes appears. The Fermi 
energy lies in the first gap for an electron density of 2 electrons per unit cell 
and it can only be in the second gap for an electron density of 8 electrons per unit 
cell.

The first gap is always indirect with the maximum of the valence band occurring at 
R. The minimum in the conduction band for the first gap occurs at X. The second 
band is sometimes direct and sometimes indirect. The valence band maximum of the 
second gap is always at M. When the second band gap is direct, the conduction band 
minimum is also at M (see Fig. \ref{fig:6}) but when the second band gap is 
indirect, the conduction band minimum is at $\Gamma$ (see Fig. \ref{fig:7}). 

\begin{figure*}
\centering
\includegraphics{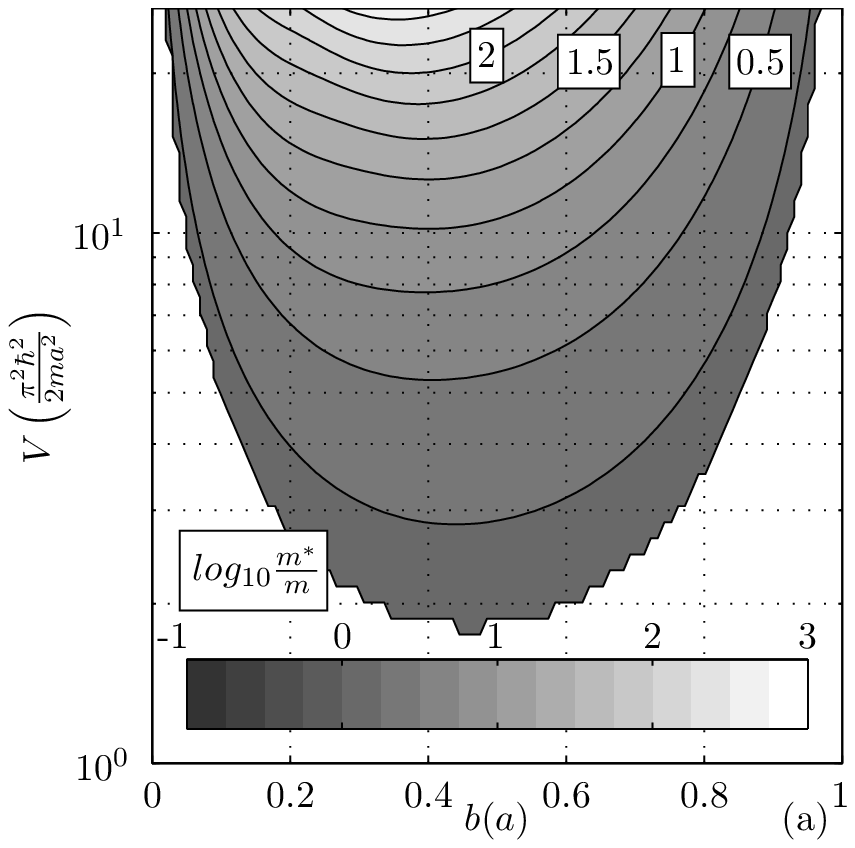}\hfill
\includegraphics{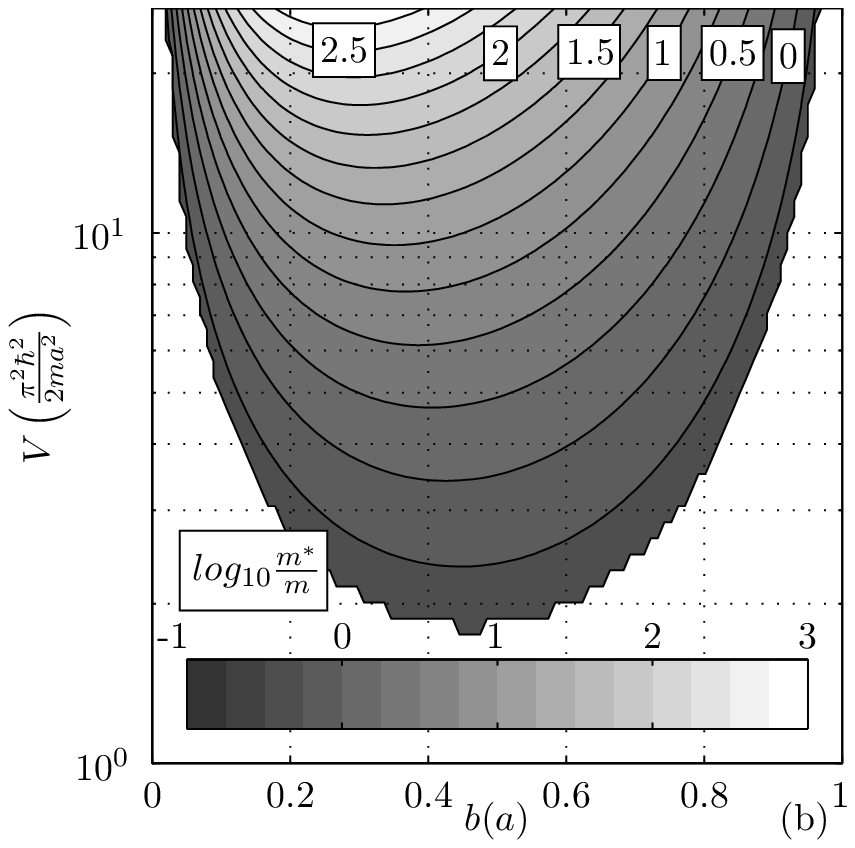}\hfill
\includegraphics{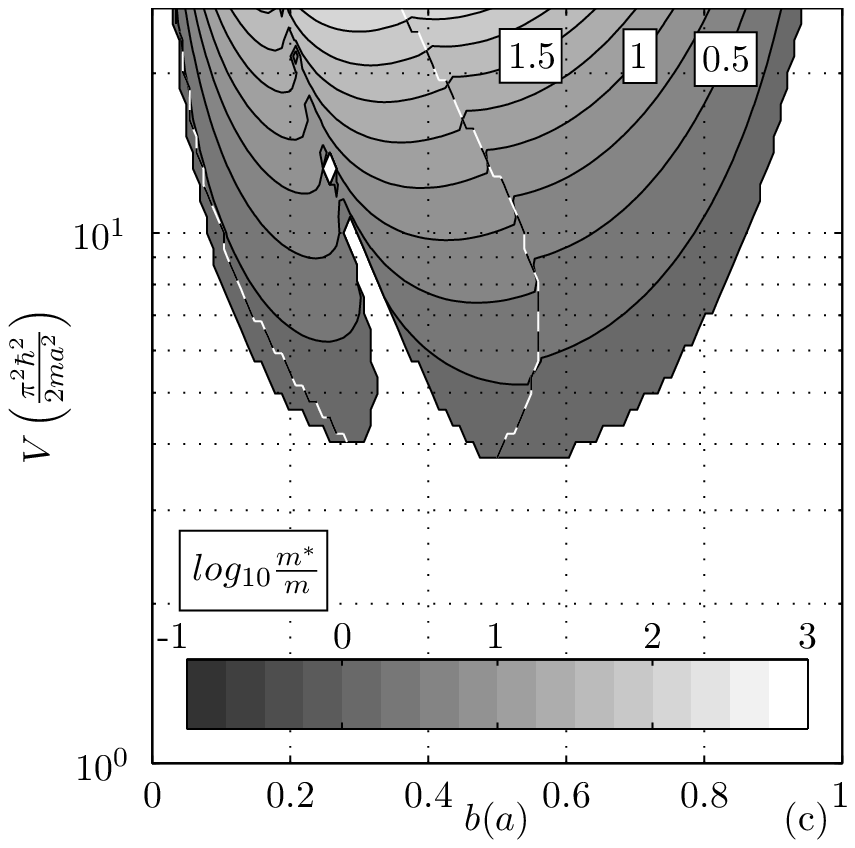}\hfill
\includegraphics{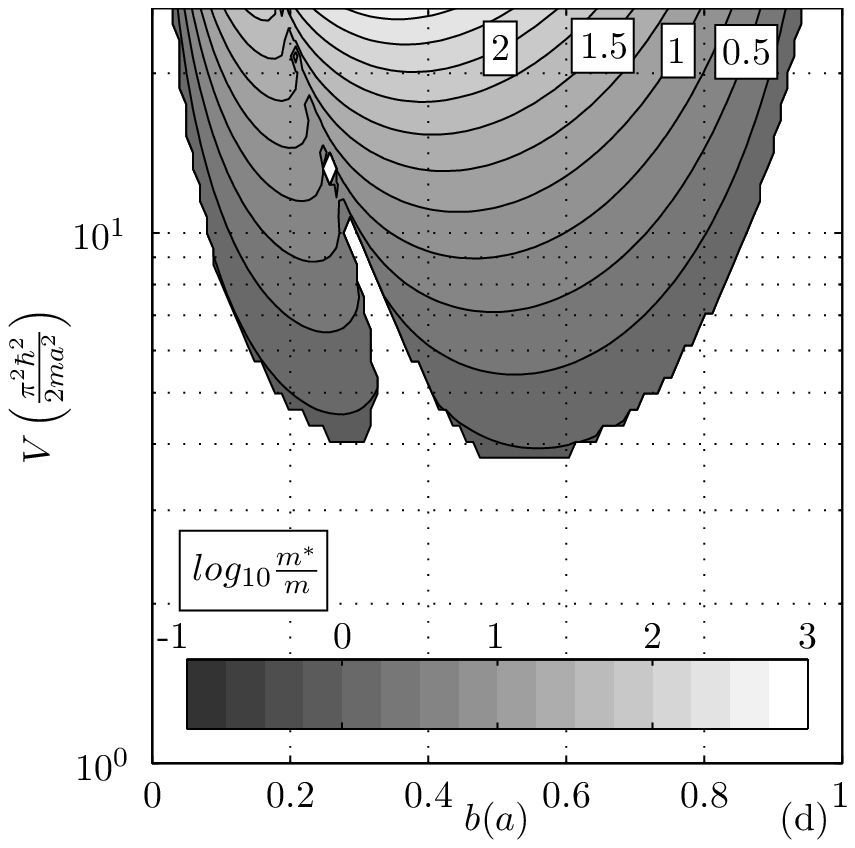}
\caption{\label{fig:10} (a) Effective masses of electrons in the first band gap. (b) 
Effective masses of holes in the first band gap. (c) The effective masses of electron 
in the second band gap. The dashed lines indicate the transitions from direct to 
indirect band gaps. The indirect band gaps are in between the two dashed lines. 
(d) The effective masses of holes in the second band gap.}
\end{figure*}

The effective masses near the top of the valence band and bottom of the conduction 
band can be found by linearizing $\alpha$ given by Eq. \eqref{eq:8} near the 
band edges and inserting this into Eq. \eqref{eq:11}. The effective masses of 
electrons and holes are, 
\begin{subequations}
\label{eq:26}
 \begin{equation}
 m_e^*  = \frac{{\hbar ^2 }}{{2a^2 }}\left( {\left. { - \frac{{d\alpha }}{{dE}}} \right|_{E = E_c } } \right) \text{ and}
 \label{eq:26a}
 \end{equation}
 \begin{equation}
 m_h^*  = \frac{{\hbar ^2 }}{{2a^2 }}\left( {\left. {\frac{{d\alpha }}{{dE}}} \right|_{E = E_v } } \right).
 \label{eq:26b}
 \end{equation}
\end{subequations}
Here $E_c$ is the energy at the bottom of the conduction band and $E_v$ is the energy 
at the top of the valence band. The effective masses are plotted in Fig. \ref{fig:10}.

The thermodynamic properties of semiconductors or insulators are typically calculated 
using the Boltzmann approximation. In this approximation, the density of states 
near the Fermi energy is described by the function,
\begin{equation}
D(E) = 
\begin{cases}
\left( 2m_h^* \right)^{3/2} \frac{\sqrt{E_v - E}}{2 \pi^2 \hbar^3}  & {E < E_v }  \\
0 & {E_v  < E < E_c }  \\
\left( 2m_e^* \right)^{3/2} \frac{\sqrt{E - E_c}}{2 \pi^2 \hbar^3}  & {E_c < E }
\end{cases}
\label{eq:27}
\end{equation}

In the Boltzmann approximation it is further assumed that in the conduction band 
the Fermi function in Eqs. \eqref{eq:14},\eqref{eq:16}--\eqref{eq:18} can be replaced by 
a Boltzmann factor $F(E) \approx exp((\mu-E)/k_BT)$ while in the valence band 
$F(E) \approx 1 - exp((E-\mu)/k_BT)$. The result for the chemical potential of a 
semiconductor in this approximation is found in many textbooks\cite{kittel_introduction_2005,ashcroft_solid_1976}. 
The chemical potential of a semiconductor or insulator is a linear function of the 
temperature. 
\begin{equation}
\mu  \approx \frac{{E_v  + E_c }}{2} - \frac{3}{4}k_B T\ln \left( {\frac{{m_e^* }}{{m_h^* }}} \right)
\label{eq:28}
\end{equation}
The Boltzmann approximation can also be used to calculate the electronic contribution 
to other thermodynamic quantities in terms of the band gap and the effective masses. 
\begin{align}
u \approx u(T = 0) + \frac{{\sqrt {2\pi } }}{{2\pi ^2 \hbar ^3 }}\left( {m_e^* m_h^* } \right)^{3/4} \exp \left( {\frac{{ - E_g }}{{2k_B T}}} \right) \nonumber \\
\times \left( {k_B T} \right)^{3/2} \left( {3k_B T + E_g } \right),
\label{eq:29}
\end{align}
\begin{align}
c_v  \approx \frac{{\sqrt {2\pi } }}{{2\pi ^2 \hbar ^3 }}\left( {m_e^* m_h^* } \right)^{3/4} \exp \left( {\frac{{ - E_g }}{{2k_B T}}} \right)\left( {k_B T} \right)^{3/2} \nonumber\\
\times \left( {\frac{{15}}{2}k_B  + \frac{{3E_g }}{T} + \frac{{E_g^2 }}{{2k_B T^2 }}} \right),
\label{eq:30}
\end{align}
\begin{align}
s \approx \frac{{\sqrt {2\pi } }}{{2\pi ^2 \hbar ^3 }}\left( {m_e^* m_h^* } \right)^{3/4} \exp \left( {\frac{{ - E_g }}{{2k_B T}}} \right)\left( {k_B T} \right)^{3/2} \nonumber\\
\times \left( {5k_B  + \frac{{E_g }}{T}} \right),
\label{eq:31}
\end{align}
\begin{align}
f \approx u(T = 0) - \frac{{\sqrt {2\pi } }}{{\pi ^2 \hbar ^3 }}\left( {m_e^* m_h^* } \right)^{3/4} \left( {k_B T} \right)^{5/2} \nonumber\\
\times \exp \left( {\frac{{ - E_g }}{{2k_B T}}} \right).
\label{eq:32}
\end{align}
Here $E_g$ is the band gap. The Boltzmann approximation assumes that there are no 
other kinks in the density of states besides the square root behavior described by 
Eq. \eqref{eq:27} within $k_BT$ of the band edge. This approximation works best 
for small band gaps. For large band gaps the bands are narrow and there are Van 
Hove singularities near the band edges. The electronic contributions to the thermodynamic 
quantities of semiconductors and insulators are exponentially suppressed at low 
temperatures by the factor $\exp(-E_g/2k_BT)$ and are often simply ignored. Equations 
\eqref{eq:29}--\eqref{eq:32} can be used to estimate the temperatures where it is no longer 
reasonable to ignore the electronic contribution to the thermodynamic properties of 
semiconductors.

\section{Conclusions}

The classic problem of electrons moving in a 1-D square-wave potential was considered 
and expressions were derived for the density of states and the effective masses 
of electrons and holes. These expressions allowed us to efficiently calculate the 
band structure and electronic contribution to the thermodynamic properties in 3-D 
separable square-wave potentials. This relatively simple model produces a wide 
range of band structures including metals, semiconductors, and insulators. Plots 
were presented showing the parameters for which band gaps appear. It was observed that 
for this class of potentials, band gaps only appear when $V > 0.9 \pi^2\hbar^2/ma^2$. 
Thus the condition for the existence of band gaps depends strongly on the lattice 
constant. The density of states at the Fermi energy and its derivative as well as the 
effective masses of the electrons and holes were also calculated. This makes it 
possible to estimate the thermodynamic properties using the standard approximations 
of the Sommerfeld expansion for metals and the Boltzmann approximation for semiconductors. 
Expressions were derived for the thermodynamic properties of semiconductors in the 
Boltzmann approximation. These simple models assume that the density of states is a 
smooth function in an energy range $k_BT$ wide. For those cases where a Van Hove 
singularity causes the density of states not to be a smooth function, the thermodynamic 
properties were calculated numerically and the results are available in the supplementary 
material. 

\newpage 

\bibliography{Separable_Crystals}

\end{document}